\documentclass[prd,aps,showpacs,showkeys]{revtex4-1}
\usepackage{bm,amsfonts,latexsym,amsmath,amssymb,amsbsy,graphicx}

\newcommand{\bb}{\begin{eqnarray}}
\newcommand{\ee}{\end{eqnarray}}
\newcommand{\ba}{\begin{align}}
\newcommand{\ea}{\end{align}}

\begin{document}

\title{\bf  Planar massless fermions in  Coulomb  and
 Aharonov-Bohm potentials}
\author{V.R. Khalilov \footnote{Corresponding author}}\email{khalilov@phys.msu.ru}
\affiliation{Faculty of Physics, Moscow State University, 119991,
Moscow, Russia}
\author{K.E. Lee}
\affiliation{Faculty of Physics, Moscow State University, 119991,
Moscow, Russia}

\begin{abstract}
Solutions to the Dirac equation are constructed for a massless charged  fermion
in  Coulomb and  Aharonov--Bohm  potentials in 2+1 dimensions.
The Dirac Hamiltonian on this background is singular and needs
a one-parameter self-adjoint extension, which can be given in terms of self-adjoint
 boundary conditions.  We show that  the virtual (quasistationary) bound states
emerge in the presence of an attractive Coulomb potential
when the so-called effective charges become overcritical and
discuss a restructuring of the vacuum of the quantum electrodynamics
when  the virtual bound states emerge.
We derive  equations, which determine the  energies and lifetimes
of virtual bound  states, find solutions of obtained equations
for some values of parameters as well as analyze the local density of states
as a function of energy in the presence of Coulomb and  Aharonov--Bohm  potentials.
\end{abstract}


.

\maketitle

\section{Introduction}

Huge interest to different
effects in the  two-dimensional (2D) systems has appeared
recently after successful fabrication
of a monolayer graphite (graphene)(see \cite{netall}
and fine Reviews \cite{ngpng,kupgc}). The single electron dynamics in
graphene is described by a massless two-component Dirac
equation \cite{ngpng,ksn,zji,vpnn,ashkl,ifh} and so
 massless Dirac excitations in graphene \cite{gnn} can provide an
interesting realization of quantum electrodynamics in 2+1 dimensions \cite{ggvo,ggvo1}.
Since, the ``effective fine structure constant" in graphene  is large,
 there appears a new possibility to study
a strong-coupling version of the quantum
electrodynamics (QED).
The induced current in the
graphene in the field of solenoid perpendicular to the plane of a sample was found to be a
finite periodical function of the magnetic flux of solenoid \cite{jmpt}.
Coulomb impurity problems, such as the vacuum polarization and screening, in graphene were
studied in \cite{vpnn,ashkl,tmksh}.
Solutions to the Dirac
equation with an Aharonov--Bohm  potential in 2+1 dimensions were also applied
 in a study of the interaction of cosmic strings with matter \cite{aw}.
 The Dirac Hamiltonians for the above problems
are essentially singular and so the supplementary definition is required
in order for they to be treated as  self-adjoint quantum-mechanical operators;
it is necessary to indicate the Hamiltonian domain in the Hilbert space
of square-integrable functions.

 An important  example of a singular Dirac Hamiltonian
is the one in a strong Coulomb field of a point-like charge described by
$4$ - potential: $A^0(r)=a/e_0r, {\bf A}=0, a>0, e_0>0$ (where $-e_0$ is
the electron charge).
We remind that the lowest bound state energy $E=m\sqrt{1-a^2}$
($m$  is the electron mass) becomes purely imaginary for $a>1$,
which implies that its interpretation as electron energy becomes
meaningless, indicates that the Hamiltonian of the system
is not a self-adjoint operator for $a>1$ and should be extended
to become a self-adjoint operator.
The latter problem are usually solving (see, fine monograph \cite{grrein})
by replacing the singular  $a/e_0r$ potential
by a  Coulomb potential cut off at small distances $R$.
In such a field, when $a$ increases, the energies of discrete states
approach the boundary of lower energy continuum, $E=-m$, and  dive into the lower continuum.
Then, discrete states turn into resonances with  finite lifetimes, which can be
described as quasistationary states with ``complex energies''.  Therefore,  an electron-positron pair is created from the vacuum:  the positron goes to infinity and the electron is coupled to the Coulomb center.
 The so-called critical charge $a_{cr}$ is determined by the condition of appearance of nonzero imaginary
part of the energy.   For  massless charged fermions in the regularized Coulomb potential, there are no discrete levels for $a<1$  due to scale invariance of the massless Dirac equation, nevertheless for $a>1$ quasistationary states emerge \cite{kupgc,ashkl,fgms,skl2,ggg,gs}.

Here we present a physically rigorous quantum-mechanical
treatment of a motion of a massless charged  fermion  in  Coulomb and  Aharonov--Bohm
potentials in 2+1 dimensions.
We stress that the presence of the AB potential  allows us to study
the influence of the particle spin on the fermion states, which is due
to the interaction between the electron spin magnetic moment
and the AB magnetic field.
This  Dirac Hamiltonian is symmetric operator so
the problem arises to construct all the self-adjoint extensions
of a given symmetric operator and then to choose  correct self-adjoint
extensions by means of physical conditions.
We construct the self-adjoint radial
 Dirac Hamiltonians on the above background
 by  the asymmetry form method \cite{vgt}
originated from von Neumann theory of self-adjoint extensions.

\section{Solutions of the radial Dirac Hamiltonian}

The space of particle quantum states in two spatial dimensions  is
the Hilbert space $\mathfrak H=L^2(\mathbb R^2)$ of square-integrable
functions $\Psi({\bf r}), {\bf r}=(x,y)$ with the scalar product
\bb
(\Psi_1,\Psi_2)=\int \Psi_1^{\dagger}({\bf r})\Psi_2({\bf r})d{\bf r},\quad d{\bf r}=dxdy.
\label{scpr}
\ee
The Dirac Hamiltonian for a massless fermion of charge
$e=-e_0<0$ in an  ($A_{\mu}$) Aharonov--Bohm
$A_0=0$, $A_r=0$, $A_{\varphi}=B/r$, $r=\sqrt{x^2+y^2}$, $\varphi=\arctan(y/x)$
and Coulomb $A_0(r) =a/e_0r$, $A_r=0$, $A_{\varphi}=0$, $a>0$
potentials, is
\bb
 H_D=\sigma_1P_2-s\sigma_2P_1+\sigma_3U(r)-e_0A_0(r),\label{diham}
\ee
where $P_\mu = -i\partial_{\mu} - eA_{\mu}$ is the
generalized fermion momentum operator.
The Dirac $\gamma^{\mu}$-matrix algebra is known to be represented  in terms of the
two-dimensional Pauli matrices $\sigma_j$ and the parameter $s=\pm 1$
can be introduced to label two types of fermions
in accordance with the signature of the
two-dimensional Dirac matrices \cite{hoso} and  is applied to characterize two states
of the fermion spin (spin ``up" and ``down") \cite{crh,khlee}.

The Hamiltonian (\ref{diham}) should
be defined as a self-adjoint operator in the Hilbert space
of square-integrable two-spinors $\Psi({\bf r}), {\bf r}=(x,y)$
with the scalar product (\ref{scpr}).
The total angular momentum $J\equiv L_z+ s\sigma_3/2$, where $L_z\equiv
-i\partial/\partial\varphi$, commutes with $H_D$, therefore, we can consider
(\ref{diham}) separately in each eigenspace  of the operator $J$
 and the total Hilbert space is a direct orthogonal sum of subspaces of $J$.

Eigenfunctions of the Hamiltonian (\ref{diham}) are (see, \cite{hkh,khlee1})
\bb
 \Psi(t,{\bf r}) = \frac{1}{\sqrt{2\pi r}}
\left( \begin{array}{c}
f_1(r)\\
f_2(r)e^{is\varphi}
\end{array}\right)\exp(-iEt+il\varphi)~, \label{three}
\ee
where $E$ is  the fermion energy, $l$ is an integer.
The wave function $\Psi$ is an eigenfunction of the
operator $J$ with eigenvalue $j=l+s/2$ and
the doublet
\bb F=\left(
\begin{array}{c}
f_1(r)\\
f_2(r)\end{array}\right)\label{doub}\ee
satisfies the equation

\bb \check h F= EF \label{radh}\ee
with
\bb
\check h=is\sigma_2\frac{d}{dr}+\sigma_1\frac{l+\mu+s/2}{r}-\frac{a}{r},\quad \mu\equiv e_0B
\label{radh0}
\ee
Thus, the problem is reduced to that for the radial Hamiltonian $\check h$
 in the Hilbert space of  doublets
$F(r)$ square-integrable on the half-line.

In the real physical space because of the existence of  the AB magnetic field
${\bf H}=(0,\,0,\,H)=\nabla\times {\bf A}= \pi B\delta({\bf r})$  there emerges
the interaction of the fermion spin magnetic moment
with the AB magnetic field in the form   $-s eB \delta(r)/r$. The
additional (spin) singular potential will reveal itself only in
the Dirac equation squared.
The ``spin'' potential is invariant under the changes  $e\to -e, s\to -s$,
and it hence suffices to consider only the case $e=-e_0<0$
and $eB\equiv -\mu<0$. Then, the potential is attractive for $s=-1$ and repulsive for $s=1$.
The influence of this singular potential
on the behavior of solutions at the origin, in fact,  is taken
into account by means of boundary conditions.

An operator, associated with the so-called differential expression $\check h$,
we shall denote by $h$.  Let  $\mathfrak H=\mathfrak L^2(0,\infty)$ be the Hilbert space of  doublets $F(r)$, $G(r)$  with  the scalar product
$$
(F,G)=\int\limits_{0}^{\infty} F^{\dagger}(r)G(r)dr=\int\limits_{0}^{\infty} [\bar f_1(r)g_1(r)+\bar f_2(r)g_2(r)]dr,
$$
so that
$\mathfrak L^2(0,\infty)=L^2(0,\infty)\oplus L^2(0,\infty)$. Here the symbol $\oplus$ denotes the direct sum.
Let us just define the operator $h^0$ in the Hilbert space $\mathfrak L^2(0,\infty)$
$$
h^0{:}\left\{\begin{array}{l}
 D(h^0)={\mathfrak D}(0,\infty), \\
 h^0F(r)=\check h F(r),
\end{array}\right.
$$
where ${\mathfrak D}(0,\infty)=D(0,\infty)\oplus D(0,\infty)$, $D(0,\infty)$ is the standard space of
smooth functions on $(0,\infty)$ with the compact support
$$
D(0,\infty)={f(r): f(r)\in C^{\infty},\;{\rm supp} f \subset [c,d],\;0<c<d<\infty}.
$$
This allows us to avoid the problems related to $r\to\infty$.

The operator  $h$ is symmetric if for any $F(r)$ and $G(r)$
\bb
\int\limits_{0}^{\infty} G^{\dagger}(r)h F(r)r dr =
 \int\limits_{0}^{\infty}[h G(r)]^{\dagger}F(r)r dr.
\label{sym}\ee
We see that $h^0$ is the symmetric operator.
Let $h$ be the self-adjoint extension $h^0$ in $\mathfrak L^2(0,\infty)$ and
consider the adjoint operator  $h^*$  (\ref{radh0})
defined by
\bb
h^*{:}\left\{\begin{array}{l}
 D(h^*)=\left\{\begin{array}{l}
F(r):F(r)\;\mbox{is absolutely continuous in}(0,\infty),\\
F,\; \check h F=G \in{\mathfrak L}^2(0,\infty),
\end{array}\right.\\
h^*F(r)=\check hF(r),
\nonumber
\end{array}\right.
\ee
i.e. $D(h^0)\subset D(h^*)$.
Since the coefficient functions of  (\ref{radh0})
are real, the deficiency indices of the operator $h^0$ are equal
so that the self-adjoint extensions of $h^0$ exist at any values of parameters
$a, \mu$, and for each $l$.
A symmetric operator $h$ is self-adjoint, if its domain $D(h)$
coincides with that of its adjoint operator  $D(h^*)$.

Integrating (\ref{sym}) by parts and taking into account  that for any doublet $F(r)$ of $D(h^*)$
$\lim\limits_{r\to \infty} F(r)=0$, Eq.  (\ref{sym}) is reduced to
\bb
\lim_{r\to 0} G^{\dagger}(r)i\sigma_2 F(r)=0. \label{bounsym}
\ee

If (\ref{bounsym}) is satisfied for any doublets from $D(h^*)\equiv D^*$
then the operator  $h^*$ is symmetric and, so,  self-adjoint.
This means that  the operator $h^0$ is essentially self-adjoint, i.e.,
its unique self-adjoint extension is its closure $h=\bar h^0$,
which coincides with the adjoint operator $h=h^*=h^{\dagger}$.
If (\ref{bounsym}) is not satisfied then the self-adjoint operator
$h=h^{\dagger}$ can be found as the narrowing  of  $h^*$ on the so-called
maximum domain $D(h)\subset D(h^*)$ \cite{vgt}.

The needed solution of (\ref{radh}) is
\begin{align}
F&=e^{-x/2}r^{\gamma_s}A'\left[v_+\Phi(a^s,\;c_s\;;x)+v_-  m_s\Phi(a^s+s,\;c_s\;;x)\right]
\nonumber \\
&\equiv AY(r,\gamma_s,E).\label{gensol}
\end{align}
Here $A'$, $A$ are constants, $x = -2i|E|r$, $a^s=\gamma_s+(1-s)/2-ie'a$, $c_s=2\gamma_s+1$, $e'=E/|E|$,
 $\gamma_s=\pm\sqrt{(l+\mu+s/2)^2-a^2}\equiv\gamma_s^{\pm}$, $m_s=(s\gamma-ie'a)/\nu, \nu=l+\mu+s/2$,
\bb
v_+=
\left(\begin{array}{c}

	1 \\ -ie'
\end{array}\right),\quad
v_-=
\left(\begin{array}{c}
	1 \\ ie'
\end{array}\right),
\ee
$\Phi(a, c; x)$ is
the confluent hypergeometric function \cite{GR}.

We denote $\gamma_s^+=\sqrt{\nu^2-a^2}\equiv \gamma$ for $a^2\leq \nu^2$ and $\gamma_s^+=i\sqrt{a^2-\nu^2}\equiv i\sigma$ for $a^2>\nu^2$. Then, for $\gamma\ne n/2$, $n=1, 2 ,\ldots$, needed  linear independent solutions are:
\begin{align}
U_1(r;E)&=Y(r,\gamma_s,E)|_{\gamma_s=\gamma},
\nonumber \\
U_2(r;E)&=Y(r,\gamma_s,E)|_{\gamma_s=-\gamma}
\label{1e35}
\end{align}
with the asymptotic behavior at $r\to 0$
\begin{align}
U_1(r;E)&=r^{\gamma}u_+{+}O(r^{\gamma+1}),
\nonumber \\
 U_2(r;E)&=r^{-\gamma}u_-{+}O(r^{-\gamma+1})
\label{e36}
\end{align}
as well as
\bb
V_1(r;E)=U_1(r;E)+\frac{a}{2s\gamma}\omega(E)U_2(r;E),
\label{e40}
\ee
where $\omega(E)={\rm Wr}(U_1,V_1)$ is the Wronskian:
\begin{align}
\omega(E)=&\frac{\Gamma(2\gamma)\Gamma\left(-\gamma+(1-s)/2-ia\right)}{\Gamma(-2\gamma)
\Gamma\left(\gamma+(1-s)/2-ia\right)}\times
\nonumber \\
&\times (-2iE)^{-2\gamma}\frac{\nu+ia+s\gamma}{\nu+ia-s\gamma}\frac{2s\gamma}{a}.
\label{wrm0}
\end{align}

The domain of the operator
$h=h^{\dagger}$ is found as the narrowing  of  $h^*$ on the
domain $D(h)\subset D^*$, so any doublet of $D(h)$ must satisfy the boundary condition (\ref{bounsym})
\bb
 (F^{\dagger}(r)i\sigma_2 F(r))|_{r=0}= (\bar f_1f_2-\bar f_2f_1)|_{r=0} =0. \label{bounsym1}
 \ee
Let us write $q=\sqrt{\nu^2-\gamma^2}$ and  $q_u=\sqrt{\nu^2-1/4}\Leftrightarrow\gamma=1/2,
\quad q_c=\nu \Leftrightarrow\gamma=0$. The quantity $q$ as a function of $l, a, \mu, s$ plays a role of
the effective charge and  $q_c$ is called the critical charge, which is
affected by the magnetic flux and the particle spin.

\section{Subcritical range ($q<q_c$). Self-adjoint boundary conditions}

By means of solutions  $U_1(r)$ and $U_2(r)$  any doublet of $D^*$  can be represented in the form (see, \cite{vgt})
\bb
F(r)=c_1U_1(r)+c_2U_2(r)+I_1(r)+I_2(r),
\label{asfr}
\ee
where $c_1$ and $c_2$ - are some constants and $I_1(r)$, $I_2(r)$  are determined by
integrals over $y$ of the  tensor product  $[U_1(r)\otimes{U}_2(y)]$.
Asymptotic behavior of $F(r)$ at $r\to 0$  essentially depends on  $\gamma$.

For $\gamma>0$ ($q<q_u$), $I_1(r)$ and $I_2(r)$ are \cite{khlee1}
\bb
I_1(r)=O(r^{1/2}),\quad I_2(r)=O(r^{1/2}),\quad r\rightarrow{0}.
\ee
It follows that  $F(r)\in{\mathfrak L}^2(0,\infty)$ implies $c_2=0$
\bb
F(r)=c_1U_1(r)+I_1(r)+I_2(r)=O(r^{1/2})\rightarrow{0},\; r\rightarrow{0}.
\ee
Then $F{\in} D^*$ and Eq. (\ref{bounsym1}) is satisfied
for $q\leq q_u$, $\gamma\geq 1/2$, which means that the initial
symmetric operator $h$  is essentially self-adjoint
and its unique self-adjoint extension is  $h=h^{\dagger}$.
Its domain  $D(h)$ is the space of
absolutely continuous doublets $F(r)$  regular at $r=0$ with $hF(r)$ belonging to $\mathfrak L^2(0,\infty)$.

For $0<\gamma<1/2$ ($q_u<q<q_c$) the left-hand side of (\ref{bounsym1}) is
$(\bar f_1f_2-\bar f_2f_1)|_{r=0} = (2s\gamma/a)(\bar{c}_1c_2-\bar{c}_2c_1)$,
or, by means of the linear transformation $c_{1,2}\rightarrow{c_\pm}=c_1\pm{ic_2}$,
is reduced to $(\bar f_1f_2-\bar f_2f_1)|_{r=0} = -i(s\gamma/a)(|c_+|^2-|c_{-}|^2)$.
Hence,  the operator $h^*$ is not symmetric and we need to construct the nontrivial self-adjoint extensions of $h^0$.  Equation (\ref{bounsym1}) will be satisfied for any $c_-$ related to $c_+$ by
$c_-=e^{i\theta}c_+$ and $0\leq\theta\leq{2\pi}$, $0\thicksim{2\pi}$.
The angle $\theta$ parameterizes the self-adjoint extensions $h_{\theta}$ of $h^0$. These extensions  vary for different $\theta$ except for two equivalent cases $\theta=0$ and $\theta=2\pi$. We denote $\xi=\tan(\theta/2)$, then
$c_2=-\xi{c_1}$, $-\infty\leq\xi\leq+\infty$, ${-\infty}\thicksim{+\infty}$.

Hence, in the range $0<\gamma<1/2$ there is one-parameter $U(1)$-family of the operators $h_{\theta}\equiv h_{\xi}$ with the domain  $D_{\xi}$
\bb
h_\xi{:}\left\{\begin{array}{l}
 D_\xi=\left\{\begin{array}{l}
F(r):F(r)\;\mbox{is absolutely continuous in}[0,\infty),
\nonumber \\
F,\check hF\in{\mathfrak L}^2(0,\infty), \\
F(r)=c[r^\gamma{u_+}-\xi r^{-\gamma}u_{-}]+O(r^{1/2}),\;|\xi|<\infty,\\
F(r)=cr^{-\gamma}u_-{+}O(r^{1/2}),\;r\rightarrow{0},\;\xi=\infty,
\end{array}\right.\\
h_\xi F=\check hF,
\nonumber 
\end{array}\right.
\ee
where $c$ is arbitrary constant.
The  operator
$h^0$ is not determined as an unique self-adjoint operator and so the additional specification of its domain, given with the real parameter $\xi$, is required in terms of the self-adjoint boundary conditions. Physically, the self-adjoint boundary conditions  show that the probability current density  is equal to zero at the origin.

The spectrum of the radial Hamiltonian is
determined by the equation  (see \cite{vgt,khlee1})
\bb
\frac{d\sigma(E)}{dE}=\frac{1}{\pi}\lim\limits_{\epsilon\rightarrow{0}}{\rm Im}\frac{1}{\omega_{\xi}(E+i\epsilon)},
\label{specfun}\ee
where the generalized function $\omega_{\xi}(E+i\epsilon)$ is obtained by the analytic continuation of the corresponding Wronskian in the complex plane of $E$; on the real axis of $E$ it is just the function $\omega(E)$ determined by (\ref{wrm0}) for $\xi=0$. We note that Eq. (\ref{wrm0})
is obtained from the corresponding Wronskian for a fermion of mass $m>0$ in the limit $m\to 0$.  Then,  the Wronskians involve the variable $\lambda=\sqrt{m^2-E^2}$  and are characterized by two cuts $(-\infty, -m]$ and $[m,\infty)$ in the complex plane of $E$, which allows us to determine the first (physical) sheet  (${\rm Re}\lambda>0$) and the second (unphysical) sheet (${\rm Re}\lambda<0$).

For $0<\gamma<1/2$ the doublet $U_\xi(r;E)$ should be chosen
in the form
\bb
U_\xi(r;E)=U_1(r;E)-\xi U_2(r;E)
\label{doubsub}
\ee
 with asymptotic behavior at $r\to 0$
 $U_\xi(r;E)=r^{\gamma}u_+-\xi r^{-\gamma}u_{-}+O(r^{-\gamma+1})$.
Solution $V_1$ is now  $V_1(r;E)\equiv V_\xi=U_\xi(r;E)+[a/2s\gamma]\omega_\xi(E)U_2(r;E)$
with $\omega_\xi(E)={\rm Wr}(U_\xi,V_\xi)=\omega(E)+2s\gamma\xi/a$
and $\omega(E)$ determined by (\ref{wrm0}).  So  $\omega_{\xi}(E)=\lim\limits_{\epsilon\rightarrow{0}}{\omega}_{\xi}(E+i\epsilon)$
and, thus, the spectral function is determined by the generalized function $F(E)=\lim\limits_{\epsilon\rightarrow{0}}{\omega}_{\xi}^{-1}(E+i\epsilon)$.
At the points, at which the function
${\omega}_{\xi}(E)=\lim\limits_{\epsilon\rightarrow{0}}{\omega}_{\xi}(E+i\epsilon)$
is not equal zero $F(E)=1/\omega_{\xi}(E)$.
It can be easily verified that the functions $\omega(E)$  and $\omega_{\xi}(E)$ are
continuous, complex-valued and  not equal to zero for real $E$;
the spectral function $\sigma(E)$ exists and is absolutely continuous.
Thus, the energy spectrum is continuous and the quantum
system under discussion does not have bound states. Bound states
 would exist if  $\omega_{\xi}(E)$ were real and the energy spectrum was determined by $\omega_{\xi}(E)=0$. One knows that real bound states (if they exist) are situated on the physical sheet of $\lambda$.

We shall suppose  that the virtual bound (quasistationary) states ``exist'' on the unphysical sheet if
their ``energies'' are determined by roots of equation $\omega_{\xi}(E)=0$.
For  $0<\gamma<1/2$, one can obtain for the real part of ${\rm Re}\omega_{\xi}(E)=0$
\bb
E=\frac{e'}2\left[\frac{\Gamma(1+2\gamma)|\Gamma(-\gamma-ia)|}{|\xi|\Gamma(1-2\gamma)|\Gamma(\gamma-ia)|}
\sqrt\frac{\nu+s\gamma}{\nu-s\gamma}\right]^{1/2\gamma}
\label{boundri10sol}\ee
and the following equation for ${\rm Im}\omega_{\xi}(E)=0$
\bb
 \pi\left(e'\gamma-\frac12\right)-\frac{3+s}{4}\arctan\frac{4a\gamma}{4\gamma^2-(1+\nu^2)(1-s)}+
 \nonumber \\
 +\sum\limits_{n=1}^{\infty}\arctan\frac{8a\gamma}{(2n+1-s)^2+4(a^2-\gamma^2)}=(p-1)\frac{\pi}{2}.
\label{boundri20}\ee
Here $p=\xi/|\xi|=\pm 1$, $p=1 (-1)$ for $\infty>\xi\geq 0 (0\geq \xi>-\infty)$.
 It can be verified that
for  $0<\gamma<1/2$ equation (\ref{boundri10sol}) does not have real root
for the values $a$, $\nu$, at which Eq. (\ref{boundri20}) is satisfied.

For definiteness, we shall put $\mu>0$. The case  $\mu<0$  can be discussed similarly with the signs of $l$ and $s$ flipped: it is just the mirror image of the case with  $\mu>0$ with respective to the $xy$-plane.
The energy range near $|E|=0$ is of interest. For  $\gamma\to 1/2$
\bb
E=e'\frac{1-2\gamma}{2|\xi|}\frac{|\Gamma(-1/2-ia)|}{|\Gamma(1/2-ia)|}
\sqrt\frac{\nu+s/2}{\nu-s/2},
\label{newb}
\ee
hence $|E|=0$  and  (\ref{boundri20}) is satisfied
by $\gamma=1/2$ for $e'=1, p=1 (\pi\geq \theta\geq 0)$ and for $e'=-1, p=-1 (2\pi\geq \theta\geq \pi)$ only if  $a^2=\nu^2-1/4$. There is the particle-hole symmetry in free particle case ($a$, $\mu=0$).

For $\gamma\to 0$, $|E|$ tends to $0$  as
$2E\approx e'(1/|\xi|)^{1/2\gamma}$ and (\ref{boundri20})
 is satisfied by $e'=\pm 1$, $\gamma=0$ only for $p=-1(0\geq\xi>-\infty,\;2\pi>\theta\geq\pi)$.
This means that the fermion states heap up close to the point $E=0$ for $E>0$ and,
conversely, for $E<0$ only when $|\xi|>1$ (see, also, \cite{vpnn})
but no fermion states will cross it  as well as no virtual bound states exist while $q<q_c$.

\section{Virtual bound (quasistationary)  states}

In the overcritical range $q>q_c (\gamma=i\sigma)$  the left-hand side of (\ref{bounsym1}) is
$$
(\bar f_1f_2-\bar f_2f_1)|_{r=0} = -(2is\sigma/a)(|c_1|^2-|c_2|^2).
$$
Thus, there is one-parameter family of the operators $h_{\theta}$ given by
\bb
h_\theta{:}\left\{\begin{array}{l}
 D_\theta=\left\{\begin{array}{l}
F(r):F(r)\;\mbox{is absolutely continuous in}[0,\infty),
\nonumber \\
F,\check hF\in{\mathfrak L}^2(0,\infty), \\
F(r)=c[e^{i\theta}r^{i\sigma} u_+ +e^{-i\theta}r^{-i\sigma} u_-]+O(r^{1/2}),\\
{r\rightarrow{0}},\quad 0 \le \theta\le \pi,\quad 0\thicksim \pi,
\end{array}\right.\\
h_\theta F=\check hF,
\nonumber
\end{array}\right.
\ee
where $c$ is arbitrary constant. We have taken into account that  $c_2=e^{i\theta}c_1$, $0\leq \theta\leq 2\pi$   is equivalent to $c_1=e^{i\theta}c$, $c_2=e^{-i\theta}c$, $0\leq \theta\leq \pi$ with
replacement $\theta\to 2\pi-2\theta$. For $\gamma=i\sigma$ the doublets $U_\theta(r;E)$ and $V_{\theta}(r;E)$ should be chosen in the form
\begin{align}
&U_\theta(r;E)=e^{i\theta}U_1(r;E)+e^{-i\theta}U_2(r;E),\label{e75}\\
&V_\theta(r;E)=U_\theta(r;E)+
\frac{ia}{4s\sigma}\omega_\theta(E)[e^{i\theta}U_1(r;E)-e^{-i\theta}U_2(r;E)],
\nonumber
\end{align}
where $U_1(r;E)$, $U_2(r;E)$ are determined by (\ref{1e35}) with $\gamma=i\sigma$,
the Wronskian is
$$
\omega_{\theta}(E)={\rm Wr}(U_{\theta},V_{\theta})=
-\frac{4is\sigma}{a}\frac{1-\tilde{\omega}(E)e^{2i\theta}}{1+\tilde{\omega}(E)e^{2i\theta}},
\quad \tilde{\omega}(E)=\frac{a}{2si\sigma}\omega(E)
$$
and $\omega(E)$ is given by (\ref{wrm0}) with $\gamma=i\sigma$.
One can verify again that  $\omega_{\theta}(E)$ are
continuous, complex-valued and is not equal to zero  for real $E$, so no bound
 states exist. Physically, this is because
 there is no natural length scale in the problem to characterize
bound states. Nevertheless,
 the  virtual (resonant) bound states can emerge when $q>q_c$;  their complex ``energies'' $E=|E|e^{i\tau}$  are determined by:
\bb
\frac{|\Gamma((1-s)/2-i(a+\sigma))|}{|\Gamma((1-s)/2-i(a-\sigma))|}
\sqrt\frac{a+s\sigma}{a-s\sigma}e^{-\pi\sigma+2\sigma\tau}=1
\label{boundsig10}\ee
and equation for the energy spectrum
\begin{align}
 &2\sigma\ln(|E|/E_0)= 2\theta -\pi\left(1+2k\right) -2\sigma{\cal C}+\arctan\frac{s\sigma}{\nu}+
 \nonumber \\
 &+\sum\limits_{n=1}^{\infty}\left(\frac{2\sigma}{n}-
 2\arctan\frac{2\sigma}{n}+\arctan\frac{2\sigma n}{n^2+\nu^2}\right).
\label{boundover}
\end{align}
where $k=0,1,2…$, a positive constant $E_0$ gives an energy scale and ${\cal C}=0.57721$ is Euler's constant. It should be emphasized that now $e'=1$ ($e'=-1$) also corresponds to the physical sheet (the unphysical sheet).

For  $\mu>0$  the fermion energies
(\ref{boundri10sol}) and (\ref{boundover})  in state with $s=-1$ ($s=1$) are less than the ones with $s=1$
($s=-1$) in the particle (hole) energy region.
This feature is due to the potential describing  the interaction of the fermion spin
magnetic moment with the AB magnetic field which is invariant under the changes  $e\to -e$, $s\to -s$.
Increasing $a$ (i.e. $\sigma$) will decrease the energy
 and increase  the number $k$. This has to do with the fact that, in reality, the so-called  Dirac
point is an accumulation point of infinitely many resonances \cite{vpnn}.

For  $\sigma\ll 1$,  Eq. (\ref{boundsig10}) has approximate  solution
$\tau \approx-(1+s)/4a+\mathrm{Im}\psi(ia)+\pi/2$,
where $\psi(z)$ is the logarithmic derivative of Gamma function  \cite{GR}
and  $\tau \approx [1+\coth(\pi/2)]\pi/2\approx (1+0.04)\pi$ for $a=1/2$, $s=1$; for $\sigma\ll 1$.
Equations (\ref{boundsig10}) and (\ref{boundover})
can be approximately satisfied near $|E|=0$ only
when $E<0$. Indeed, for $a>\nu$, $\sigma>0$
 (\ref{boundsig10}) is satisfied  only when  $e'=-1$, $\tau>\pi$
 and the right hand side of the equation (\ref{boundover})
is negative.  Then, for $\sigma\ll 1$ the energy spectrum is determined by
\begin{align}
 &E_{k,\theta,s}= E_0 \cos(\tau)
 \exp\left[-\pi(1+2k)/2\sigma+\theta/\sigma-\right.
 \nonumber \\
 &\left.-({\cal C}+(1-s)/2
 +\pi^2/6-(\pi\coth\pi a)/2a)\right].
\label{energyres}
\end{align}

These energies have an essential singular point at $\sigma=0$  \cite{ashkl,kupgc,ggg}.
The infinite number of quasistationary levels is
related to the long-range character of the Coulomb potential \cite{ggg,vpnn,ashkl}.

Therefore, the virtual bound states  abruptly emerge in the presence
of an attractive Coulomb potential at $q>q_c$.
   The imaginary part of $E_{k,\theta,s}$ define the width of
virtual resonant states or the inverse lifetimes (decay rates) due
to the interaction with the Coulomb center. It follows from
(\ref{boundsig10}) that $\sin\tau\sim 0.2\cos\tau$ (for $\sigma\ll 1$)  so
 the width of resonant states are $\sim |E_{k,\theta,s}|$, hence,
they are practically bound states.
In the overcritical range the wave functions oscillate
with  frequency $2\sigma\ln2|E|r$ as $r\to 0$, which is due to
the asymptotic behavior of function
$\Phi_{i\sigma}(a_1^s)\sim e^{2i\sigma\ln2|E|r}$ at small $x$.
  Such a situation is akin to the fall of a particle
to the field center in the nonrelativistic quantum mechanics \cite{grrein}.

In the relativistic quantum mechanics the emergence of virtual bound levels
must entail a restructuring of the vacuum. If the emergent virtual level was empty, an
electron-hole pair will be created: the electron from the filled valence band (the Dirac sea)
occupies this virtual level with diverging lifetime and shields the center,
while the emergent (in the valence band) hole is ejected to infinity.
The emergent virtual level could be occupied by an electron in the adatom \cite{nkpnp}; then,  no electron-hole pair will be created but the vacuum will be restructured.

\section{The local density of fermion states}

 The experimentally accessible quantity is
the local density of states (LDOS) as a function of distance
from the origin; the LDOS per unit area is determined by \cite{vpnn}
\bb
N(E,r)=\sum\limits_{l=-\infty}^{\infty}|\Psi(t,{\bf r})|^2=\sum\limits_{l=-\infty}^{\infty}n_l(E,r),
\quad n_l(E,r)=\frac{|f_1(r,E,l)|^2+|f_2(r,E,l)|^2}{2|A_l(E)|^2\pi r},
\label{dens}
\ee
where $f_1(r,E,l)/A_l(E)$ and $f_2(r,E,l)/A_l(E)$ are the doublets normalized
(on the half-line with measure $dr$) by imposing orthogonality on the energy scale and
$A_l(E)$ is the normalization constant.

For $\gamma\geq 1/2, q\leq q_u$ the LDOS is
determined by
\bb
N_{reg}(E,r)=\frac{e^{\pi ae'}}{2\pi^2 r}\sum\limits_{l'=-\infty}^{\infty}
\frac{(2|E|r)^{2\gamma}|\Gamma(\gamma+1+iae')|^2}{\Gamma^2(2\gamma+1)}
|\Phi_\gamma(a^s)|^2,
\label{zero}\ee
where the sum is taken over $l$ satisfying the inequality $\sqrt{(l+\mu+s/2)^2-a^2} \geq 1/2 $,  $\Phi_\gamma(a^s)\equiv \Phi(\gamma+(1-s)/2-ie'a, 2\gamma+1, x)$ and
$N_{reg}(E,r)$ is expressed through regular functions  at $r=0$.
In the limits $a=0, \mu=0$ the function  $\Phi_\gamma(a^s)$ is reduced to the Bessel functions of integer order  and the free density of states  is easily recovered from (\ref{zero})  to be  $N(E,r)=|E|/2\pi$.
We shall consider the LDOS for the (spin up) case $s=1$  and comment
the LDOS with $s=-1$ since the latter can be analyzed taking into account
 the obvious relation
\bb
\gamma(\pm l, s=1, \mu, a)=\gamma(\pm l+1, s=-1, \mu, a).
\label{grel}
\ee
For small effective charge, the LDOS at different distances $r$ from the origin  are given in FIG. 1, for $s=1$ and FIG. 2, for $s=-1$.

For $1/2>\gamma>0 (q_u<q<q_c)$, the LDOS should
be constructed by means of Eq. (\ref{doubsub}) by summing over $l'$:
\bb
N_{\xi}(E,r)= \frac{1}{2\pi r}\sum\limits_{l'}\frac{n^I_l
-4\xi|n^{II}_l|\left\{\cos\left(\mathrm{Arg}[n^{II}_l]\right)+e'\gamma\sin\left(\mathrm{Arg}[n^{II}_l]\right)/a\right\}}{A(\gamma,E)+\xi^2A(-\gamma,E)-2\xi B(\gamma,E)},
\label{quqqc}
\ee
where the sum is taken over $l$ from   $1/2>\sqrt{(l+\mu+s/2)^2-a^2}>0$,
$$
n^I_l=\frac{2\nu(\nu+s\gamma)}{a^2}r^{2\gamma}|\Phi_\gamma(a^s)|^2+\frac{2\nu(\nu-s\gamma)}{a^2}\xi^2r^{-2\gamma}|\Phi_{-\gamma}(a^s)|^2, $$
$$
n^{II}_l=\Phi_{\gamma}(a^s)\Phi^*_{-\gamma}(a^s),\quad \xi=\tan(\theta/2),\quad 0\leq\theta\leq2\pi
$$
and
$$
A(\gamma,E)=\frac{2\pi e^{-\pi ae'}\Gamma^2(2\gamma+1)\nu(\nu+s\gamma)}
{|\Gamma(\gamma+1+iae')|^2(2|E|)^{2\gamma}a^2},\\
B(\gamma,E)=\frac{\pi e^{-\pi ae'}\Gamma(2\gamma+1)\Gamma(-2\gamma+1)}{|\Gamma(\gamma+1+iae')||\Gamma(-\gamma+1+iae')|}.
$$
When $a\neq 0$ Eqs. (\ref{zero}) and (\ref{quqqc})  contain the energy sign  $e'$,  which means that the particle-hole symmetry is lost.
Writing, for example for $s=1$,  $\gamma_{l=0,-1}=\sqrt{(1/2\pm\mu)^2-a^2}$, we see that the partial terms with $l=0, -1$  give different contributions to the LDOS in the presence of the magnetic flux.
The peaks at positive energies for some $\theta$ in the subcritical range (see, FIG. 3 in which $s=1$) in which the LDOS exhibits is due to singular (at $r\to 0$) solutions
(compare with results \cite{vpnn}).  It is also seen that the attractive Coulomb potential brings locally a reduction of spectral weight in the negative energy range, the opposite happens to the positive range; the effect is strongest near the Coulomb center.
This behavior of the spectrum near the Dirac point can
be understood from an investigation of the quantized energies (\ref{energyres}).

In the overcritical range  $\gamma=i\sigma, \quad 0\geq \theta\geq \pi$ with using (\ref{e75}),  one obtains
\bb
N_{\theta}(E,r)=\frac{1}{2\pi^2 r}\sum\limits_{l'}\frac{n^I_l(x)+4\nu |n^{II}_l(x)|\cos[2\sigma\mathrm{ln}r+2\theta+\mathrm{Arg}(n^{II}_l(x))
]}{n^I_l(\infty)+4\nu |n^{II}_l(\infty)|\cos[2\theta-2\sigma\mathrm{ln}(2|E|)+g(\sigma)
]},\label{overcr}
\ee
where now $l'$ denotes the sum taken over $l$ from $a^2>(l+\mu+s/2)^2$,
$$
n^I_l(x)=\frac{(a+se'\sigma)[|\Phi_{i\sigma}(a^s)|^2+|(s\sigma-ae')/\nu|^2|\Phi_{i\sigma}(a^s+s)|^2]}{a}+ $$
$$
+\frac{(a-se'\sigma)[|\Phi_{-i\sigma}(a^s)|^2+|(s\sigma+ae')/\nu|^2|\Phi_{-i\sigma}(a^s+s)|^2]}{a},
$$
$$
n^{II}_l(x)=(\nu+is\sigma)\Phi_{i\sigma}(a^s)\Phi^*_{-i\sigma}(a^s)/a^2, \quad
n^{I,II}_l(\infty)=n^{I,II}_l(x)|_{x\to \infty},
$$
$$
g(\sigma)=\mathrm{Arg}[\Gamma^2(2i\sigma+1)/\Gamma(1+i\sigma+isae')\Gamma(1-i\sigma+isae')]
+\arctan(s\sigma/\nu).
$$
The total LDOS  is $N(E,r)=N_{reg}(E,r)+N_{\xi}(E,r)+N_{\theta}(E,r)$.


\newpage

\begin{figure}[h!]
\centering
{\includegraphics[width=4.2cm]{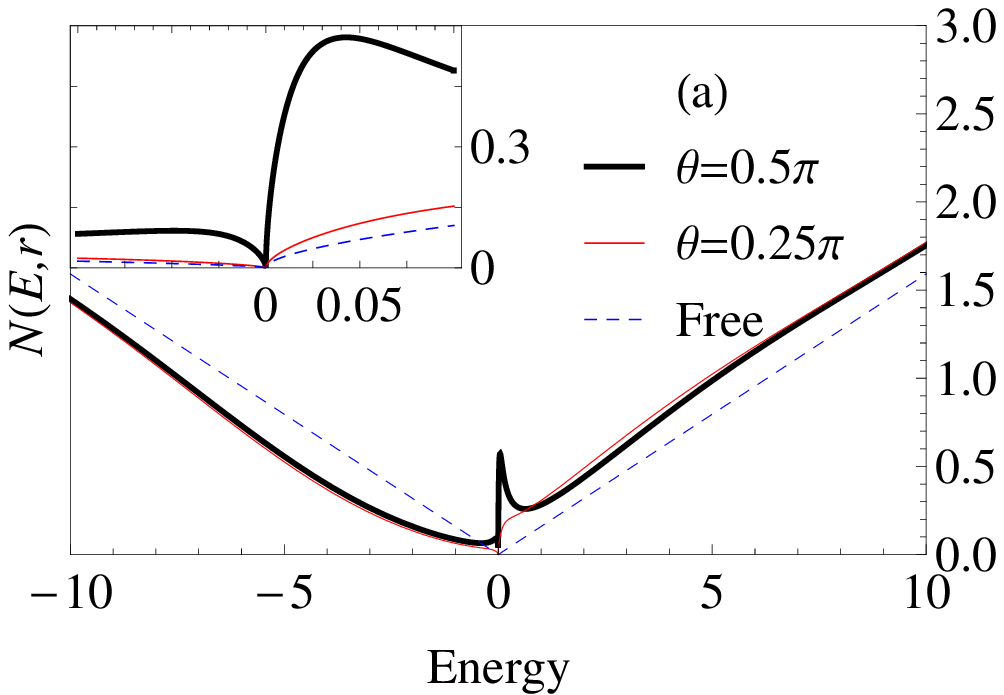}}
{\includegraphics[width=4.2cm]{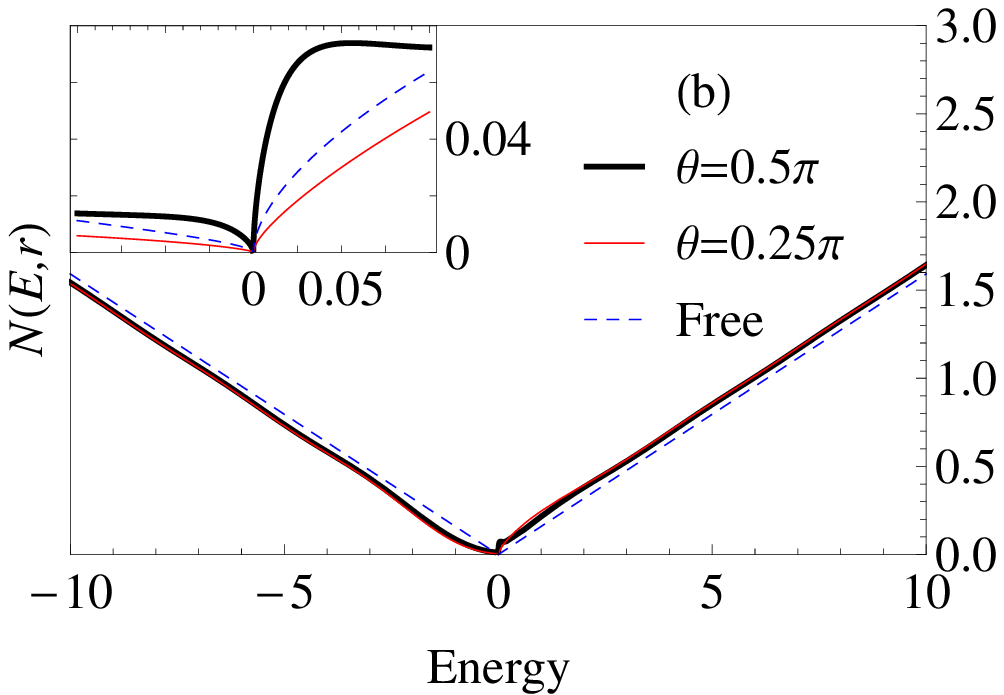}}
\caption{Total LDOS $N(E,r)=N_{reg}(E,r)+N_\xi(E,r)$ for $a=0.3,\mu=0.1,s=1$ and $r=0.3\;({\rm a}),
r=1\;({\rm b})$; the insets are  magnifications for $E\approx0$. The free DOS for $a=0,\mu=0$ is included for comparison (dashed line).} \label{fig.1}
\end{figure}

\begin{figure}[h!]
\centering
{\includegraphics[width=4.2cm]{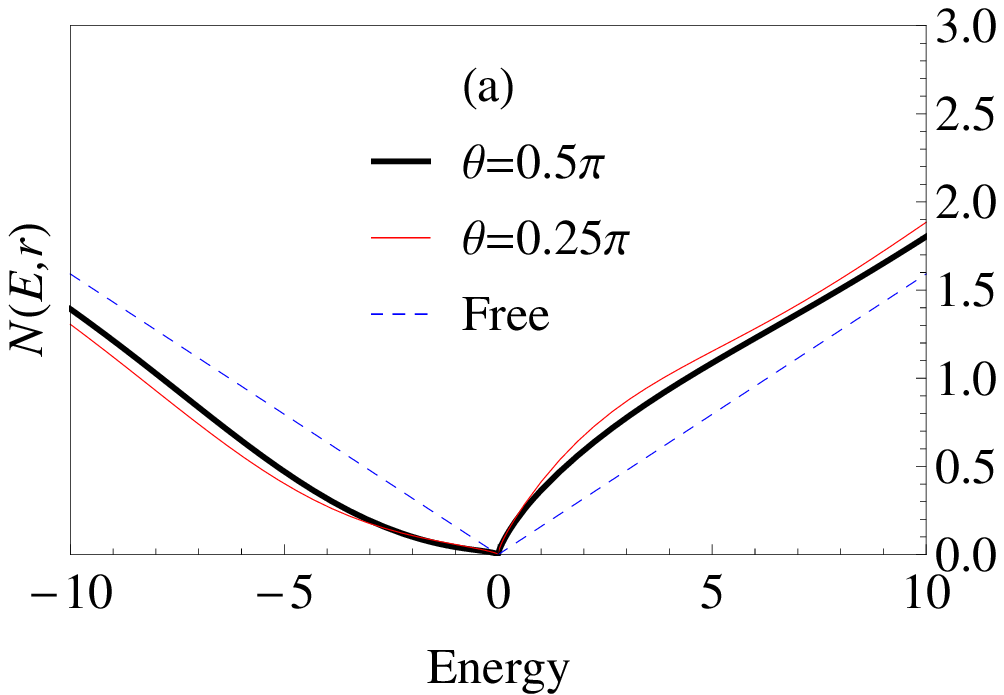}}
{\includegraphics[width=4.2cm]{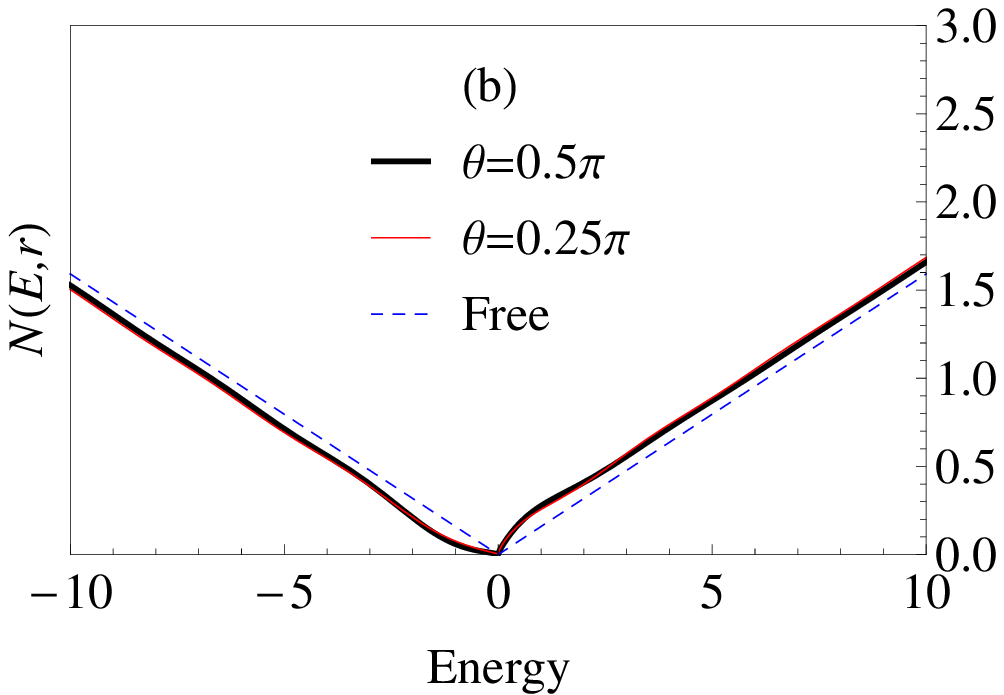}}
\caption{Total LDOS $N(E,r)=N_{reg}(E,r)+N_\xi(E,r)$ for $a=0.3,\mu=0.1, s=-1$ and $r=0.3\;({\rm a}),
r=1\;({\rm b})$ The free DOS for $a=0,\mu=0$ is included for comparison (dashed line).} \label{fig.2}
\end{figure}

\begin{figure}[h!]
\centering
{\includegraphics[width=4.2cm]{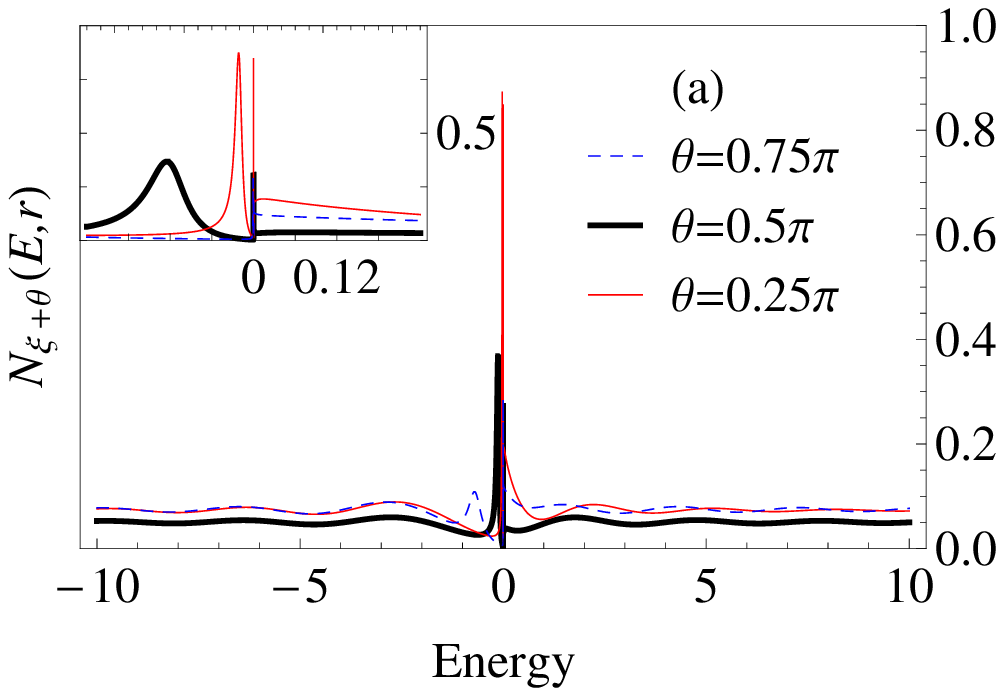}}
{\includegraphics[width=4.2cm]{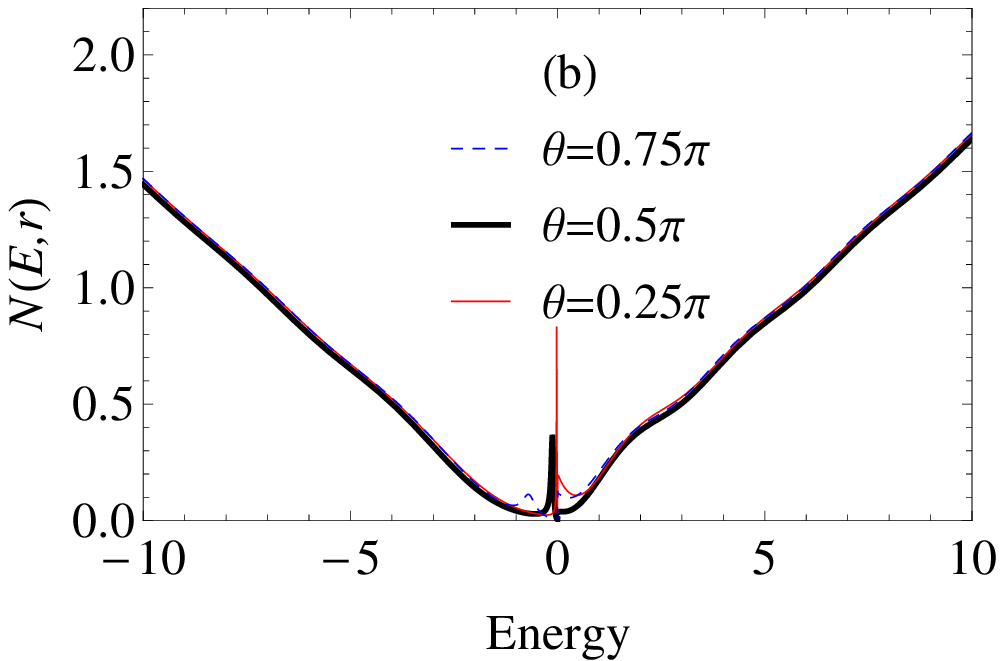}}
\caption{$N_{\xi+\theta}(E,r)=N_\xi(E,r)+N_\theta(E,r)$ with $l=0$ ($\gamma\approx0.0035$) and $l=-1$; the inset is a magnification for $E\approx 0$ $({\rm a})$. Total LDOS $N(E,r)=N_{reg}(E,r)+N_\xi(E,r)+N_\theta(E,r)$  $({\rm b})$. On all panels: $a=0.59999$, $\mu=0.1$, $r=1$.}
\label{fig.3}
\end{figure}

\begin{figure}[h!]
\centering
{\includegraphics[width=4.2cm]{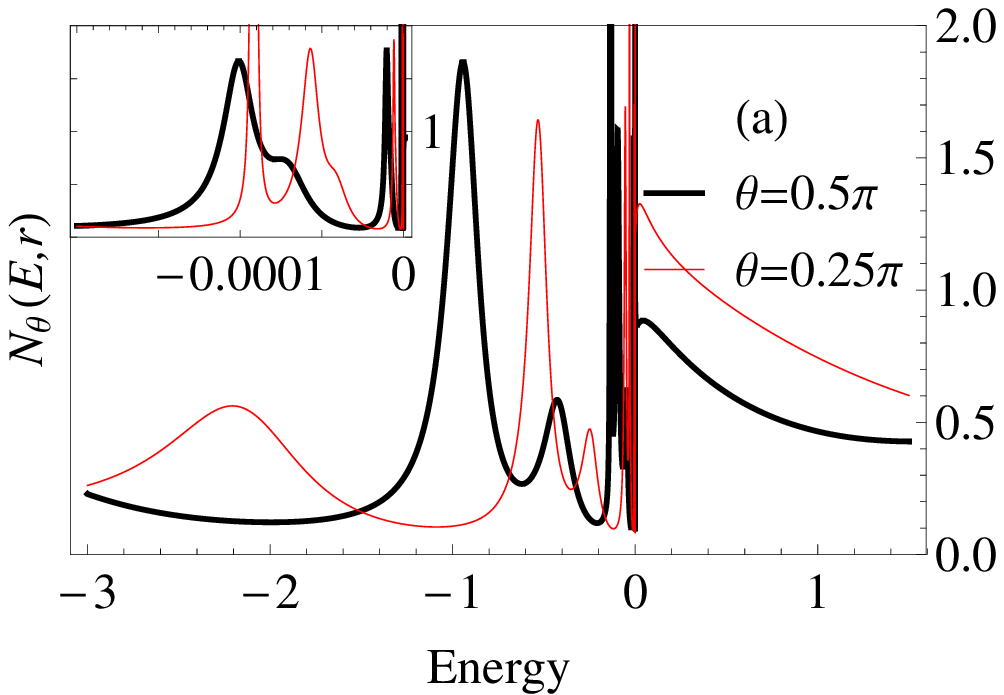}}
{\includegraphics[width=4.2cm]{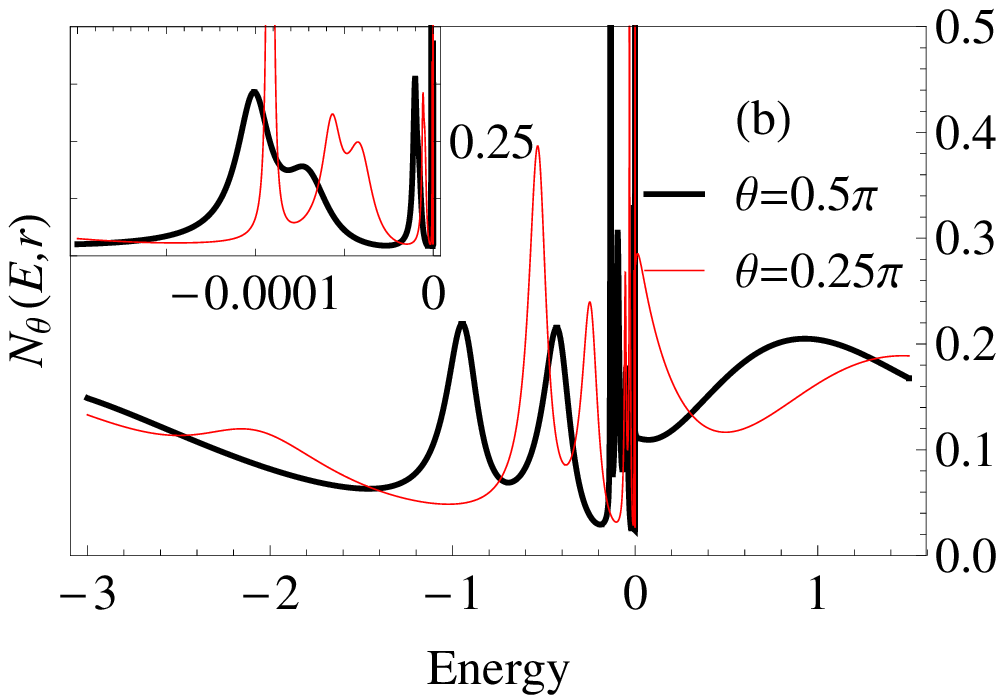}}
\caption{LDOS $N_\theta(E,r)$ with $l=-2,-1,0$ for $a=1.5,\mu=0.1$
($\sigma\approx 0.539, 1.446, 1.375$) and $r=0.3\;({\rm a}),
r=1\;({\rm b})$; the insets are magnifications for $E\approx 0$.}
\label{fig.4}
\end{figure}


\begin{figure}[h!]
\centering
{\includegraphics[width=4.2cm]{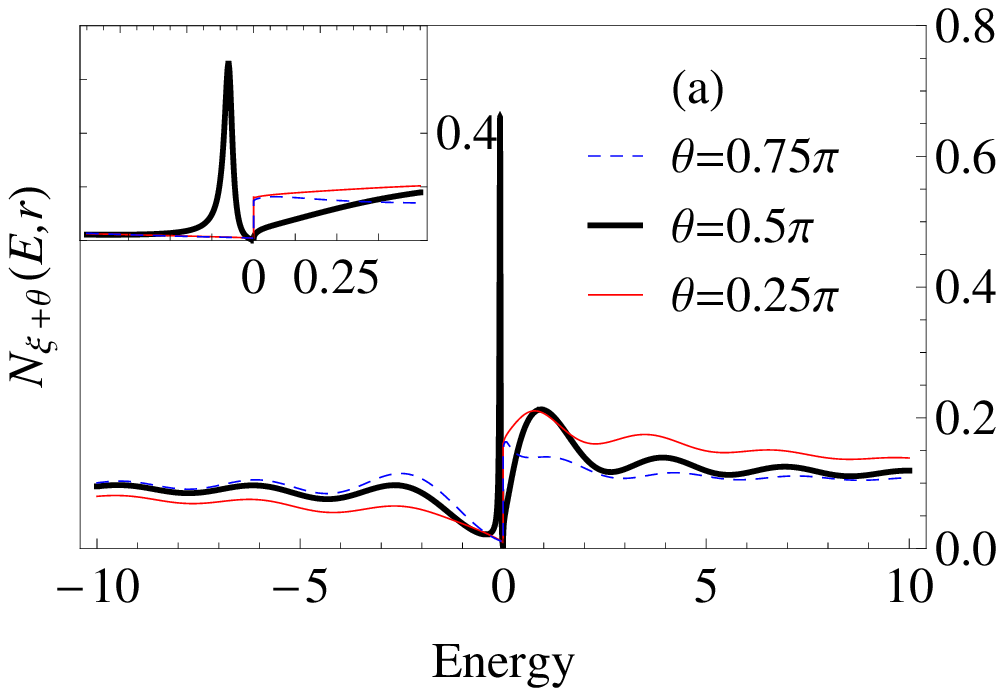}}
{\includegraphics[width=4.2cm]{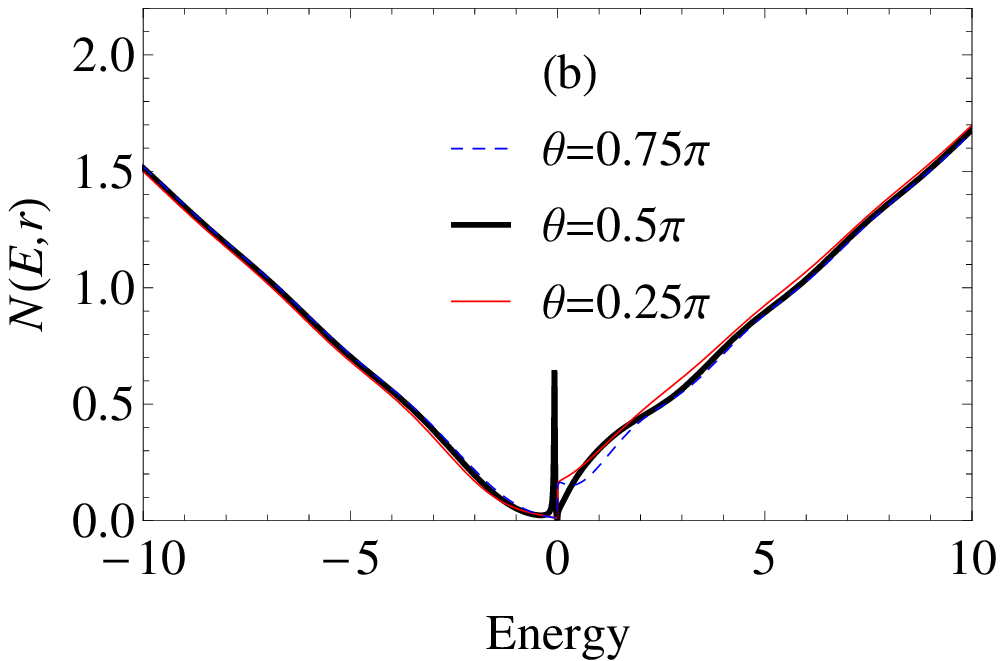}}
\caption{LDOS $N_{\xi+\theta}(E,r)=N_\xi(E,r)+N_\theta(E,r)$ for $l=0$ ($\gamma\approx0.3162$) and $l=-1$ ($\sigma\approx0.003$); the inset is a magnification for $E\approx0$ $({\rm a})$. Total LDOS $N(E,r)=N_{reg}(E,r)+N_\xi(E,r)+N_\theta(E,r)$  $({\rm b})$. On all panels $a=0.45001$, $\mu=0.05$, $r=1$.}
\label{fig.5}
\end{figure}
It should be commented: Since the summing range over $l$ for $s=-1$ is changed
as compared to the one for $s=1$, little peaks in FIG. 2 absent.
Families of the curves for the LDOS with $s=-1$ are qualitatively like
to the ones given in FIGs. 3, 4
 at the same values of $a, \mu, \xi, \theta$ except to the shift $l\to l+1$.
Importantly,  the LDOS  exhibits  resonances of the width  $\sim |E_{k,\theta,s}|$ at the negative
energies (\ref{energyres}), which decay away
from the impurity (see, FIG. 4 for $s=1$). Strong resonances
appear in the vicinity of the Dirac point and signal
the presence of the quasistationary  states while
at positive energies the LDOS exhibits periodically decaying oscillations
(see, \cite{vpnn,ashkl}).
Increasing the effective charge will cause the resonances
to migrate downwards in energy and their number to increase. This is because,
in reality, the Dirac point is an accumulation point of infinitely many resonances \cite{vpnn}.

FIG. 5 shows there is indeed  the single resonance in the hole region
when $\sigma\to 0$ at $\theta =\pi/2$ and only for $s=1$, which is in good accord with (\ref{energyres}).

It should be noted that the local and total density of states
in the pure Aharonov--Bohm potential with half-integers $\mu$ in graphene are calculated in \cite{ssladd}.
It was shown in \cite{ssladd} that: 1) the peak of the LDOS, due
to the divergent as $1/\sqrt{r}$ at the origin zero mode solution
of the Dirac equation, should be observed at the Fermi
level in graphene without gap in the quasiparticle spectrum; 2) when the energy
is increased the LDOS very quickly reduces to the free density of states. These results
can be obtained from Eqs. (\ref{zero}) and (\ref{quqqc}) putting in them $a=0, l=0, \mu=1/2$.
Exact solutions to the Dirac equation  in the pure Aharonov–Bohm
potential in 2+1 dimensions was found and discussed  in \cite{khl10}
for fermion bound states
with the particle spin taken into account.






\vspace{2cm}

\begin{thebibliography}{55}

\bibitem{netall} K. S. Novoselov et al., Science, {\bf 306}, 666 (2004).

\bibitem{ngpng} A.H. Castro Neto, F. Guinea, N.M. Peres, K.S. Novoselov,
and A.K. Geim, Rev. Mod. Phys., {\bf 81} 109 (2009).

\bibitem{kupgc} V. N. Kotov, B. Uchoa, V. M. Pereira, F. Guinea and A. H. Castro Neto, Rev. Mod. Phys. {\bf 84}, 1067 (2012).

\bibitem{ksn} K.S. Novoselov et al, Nature, {\bf 438}, 197 (2005).

\bibitem{zji} Z. Jiang, Y. Zhang, H.L. Stormer, and P. Kim,
 Phys. Rev. Lett., {\bf 99} 106802 (2007).

\bibitem{vpnn} V. M. Pereira, J.  Nilsson, and A.H. Castro Neto,
Phys. Rev. Lett., {\bf 99} 166802 (2007).

\bibitem{ashkl} A.V. Shytov, M.I. Katsnelson, and L.S. Levitov,
Phys. Rev. Lett., {\bf 99} 236801 (2007).

\bibitem{ifh} I.F. Herbut, Phys. Rev. Lett., {\bf 104} 066404 (2010).

\bibitem{gnn} A.K. Geim and K.S.Novoselov, Nat. Mater. {\bf 6}, 183 (2007).

\bibitem{ggvo} J. Gonzarlez, F. Guinea, and M.A.H. Vozmediano, Nucl.
Phys., {\bf B 424}, 595 (1994).

\bibitem{ggvo1} J. Gonzarlez, F. Guinea, and M.A.H. Vozmediano,
J. Low Temp. Phys., 99, {\bf 287} (1995).

\bibitem{jmpt} R. Jackiw, A. I. Milstein, S.-Y. Pi and I. S. Terekhov, Phys. Rev. {\bf B80}, 033413 (2009).

\bibitem{tmksh} I.S. Terekhov, A.I. Milstein, V.N. Kotov, and O.P.
Sushkov, Phys. Rev. Lett., {\bf 100} 076803 (2008).

\bibitem{aw} M.G. Alford and F. Wilczek, Phys. Rev. Lett. {\bf 62}
1071 (1989).

\bibitem{grrein} W. Greiner, J. Reinhardt, {\sl Quantum Electrodynamics},
$4^{th}$ ed. (Springer-Verlag, Berlin Heidelberg, 2009).

\bibitem{fgms}  P.I. Fomin, V.P. Gusynin, V.A. Miransky, and Yu.A. Sitenko, Riv. Nuovo Cimento, {\bf 6}, 1 (1983).
\bibitem{skl2} A.V. Shytov, M.I. Katsnelson, and L.S. Levitov, Phys. Rev. Lett., {\bf 99}, 246802 (2007).

\bibitem{ggg} O.V. Gamayun, E.V. Gorbar, and V.P. Gusynin,
Phys. Rev., {\bf B80}, 165429 (2009).

\bibitem{gs} K.S. Gupta, and S. Sen, Mod. Phys. Lett., {\bf A24}, 99 (2009).


\bibitem{vgt} B.L. Voronov, D.M. Gitman, and I.V. Tyutin,
Theoretical and Mathematical Physics, {\bf 150}, 34 (2007).




\bibitem{hoso} Y. Hosotani, Phys. Lett., {\bf B319}, 332 (1993).

\bibitem{crh} C.R. Hagen, Phys. Rev. Lett., {\bf 64}, 503 (1990).

\bibitem{khlee} V.R. Khalilov and K.-E. Lee, Journ. Phys., {\bf A44}, 205303 (2011).

\bibitem{hkh} V.R. Khalilov, Phys. Rev., {\bf A71}, 012105 (2005).


\bibitem{khlee1} V.R. Khalilov and K.-E. Lee, Mod. Phys. Lett., {\bf A26}, No 12, 865 (2011).

\bibitem{GR} I.S. Gradshteyn and I.M. Ryzhik, {\sl Table of Integrals,
 Series, and Products}, $5^{th}$ ed. (Academic Press, San Diego, 1994).



\bibitem{nkpnp} A. H. Castro Neto, V. N. Kotov, V. M. Pereira, J. Nilsson, N. M. Peres and B. Uchoa,
Solid State Commun. {\bf 149}, 1094 (2009).

\bibitem{ssladd} A.O. Slobodeniuk, S.G. Sharapov, and V.M. Loktev, Phys. Rev., {\bf B82}, 075316 (2010).
\bibitem{khl10} V.R. Khalilov, Theoretical and Mathematical Physics, {\bf 163}, 511 (2010).

\end{thebibliography}
\end{document}